\documentclass[aps,prd,amsfonts,nofootinbib,supercriptaddress]{revtex4}
\usepackage{amssymb,amsmath}
\usepackage{graphicx}
\usepackage{epsfig}
\usepackage{bbm}

\newcommand{\beq}{\begin{equation}}
\newcommand{\eeq}{\end{equation}}
\newcommand{\bea}{\begin{eqnarray}}
\newcommand{\eea}{\end{eqnarray}}
\newcommand{\vc}[1]{{\textbf{#1}}}
\newcommand{\s}[1]{{\mathsf{#1}}}

\begin{document}

\title{Decoherence from Isocurvature Perturbations in Inflation}

\author{Tomislav Prokopec}
\author{Gerasimos I.~Rigopoulos\footnote{Current address: Helsinki Institute of Physics, P.O.Box 64,
FIN-00014, University of Helsinki,
Finland
}}
\affiliation{Institute for Theoretical Physics, Utrecht University,\\
Postbus 80.195, 3508 TD Utrecht, The Netherlands}

\begin{abstract}

\vskip -3.2cm
\rightline{\hskip 3cm ITP-UU-06/51, SPIN-06/41}
\vskip 2.7cm

\noindent We discuss the possible role of isocurvature perturbations
for the quantum decoherence of the curvature perturbation during
inflation. We point out that if the inflaton trajectory in field space
is curved, the adiabatic mode is generically coupled to the isocurvature mode and thus tracing out the latter can cause the curvature perturbation to decohere. We explicitly investigate this suggestion in a model of inflation with two decoupled massive fields and find that decoherence is effective for a wide range of mass ratios. In particular, we calculate the entanglement entropy for this model and show that it grows after horizon exit, providing a quantitative measure
for decoherence.

\end{abstract}

\maketitle

\section{Introduction}

Inflation has become the leading paradigm of early universe
cosmology not least because of its ability to imprint scale
invariant inhomogeneities on superhorizon scales via a causal
mechanism \cite{inflation}. These inhomogeneities are thought to
provide the seeds which later become the temperature anisotropies
in the Cosmic Microwave Background and the Large Scale Structure
in the Universe. In inflation, the origin of these initial
perturbations lies in quantum fluctuations of matter fields,
amplified during a period of quasi-exponential expansion, which
source adiabatic and isocurvature cosmological perturbations.
The adiabatic perturbation is a variation in total energy density or
equivalently in the gravitational potential, while isocurvature
perturbations are variations in the entropy density of various
matter components. All inflationary models share this basic
mechanism, with different models characterized by small deviations
from exact scale invariance. Along with other calculable
properties, these deviations are particularly appealing as they
allow for inflationary models to be tested against increasingly
precise observations~\cite{Spergel:2006hy}.

Inflationary calculations are based on the quantum mechanics of
scalar fields in expanding spacetimes~\cite{Guth:1985ya}, where
the relevant observable is the amplitude of the field's Fourier
modes. Although treated as a quantum mechanical variable, this
amplitude is interpreted as a stochastic random variable described
by a gaussian distribution, with the variance given by the
power-spectrum. This interpretation, used in CMB analyses and
simulations of Large Scale Structure, proves to be very accurate
for calculational purposes; quantum correlation functions, exactly
calculated, are practically the same as the corresponding
correlation functions obtained by using the relevant classical
probability distribution~\cite{Polarski:1995jg}.

The consistency of this stochastic interpretation requires a density
matrix which is diagonal in the amplitude basis. However, the
density matrix of inflationary perturbations is \emph{not}
automatically diagonal in this basis (see eg. \cite{lombardo1}).
One cannot assign a specific amplitude to the Fourier modes which must
thus be considered to exist in coherent superpositions of different
amplitudes. Prior to a stochastic interpretation, an as-of-yet
undisclosed decoherence mechanism is required to diagonalize the
density matrix.

The phenomenon of decoherence is ubiquitous in quantum physics and
originates from couplings of the system of interest with degrees
of freedom belonging to some unobservable environment~\cite{Zurek,
Joos}. Various arguments and calculations suggesting that a form of such environmental
decoherence can indeed occur for inflationary perturbations have
been put forward in~\cite{Brandenberger:1990bx, calzetta-hu, lombardo-mazzitelli, matacz, campo-parentani, Burgess:2006jn,
Martineau, Kiefer:2006je, lombardo2}. In \cite{calzetta-hu, lombardo-mazzitelli, matacz, campo-parentani, Burgess:2006jn,
Martineau, lombardo2}, the environment is taken to consist of the short wavelength modes which are coupled to the long wavelength modes via non-linear couplings. In this paper, we point out
that the decoherence of cosmological perturbations can occur in a
precisely defined dynamical setting if the role of the environment
is played by unobservable isocurvature perturbations generated
during multi-field inflation. The coupling in this case are bilinear and no non-linearities are invoked.

\section{Quantum mechanics of inflationary perturbations}

To study decoherence in inflation it is convenient to use the
Schr\"{o}dinger picture which we briefly review in this section. The action for a
free massive scalar field is
    \beq
    S=-\frac{1}{2}\int d^4x \sqrt{-g}\,
     \left(\partial_\mu\phi\partial^\mu\phi+m^2\phi^2\right)
\label{action}
    \eeq
where we take the spacetime metric to be
    \beq
    ds^2=g_{\mu\nu}dx^\mu dx^\nu=-dt^2+a(t)^2\delta_{ij}dx^idx^j\,.
    \eeq
Then, the canonical momentum conjugate to the field $\phi$,
    \beq
    \pi=\frac{\partial\mathcal{L}}{\partial
    (\partial_t\phi)}=a^3\partial_t\phi\,,
    \eeq
allows us to write the Hamiltonian $H \equiv \int d^3x
\,(\pi\partial_t\phi-\mathcal{L})$
    \beq
    H=\int d^3x\,{1\over 2}\left(\frac{\pi^2}{a^3}+a(\partial_i\phi)^2 +
    a^3m^2\phi^2\right)\,.
    \eeq
Quantization dictates the replacement $\pi(\mathbf{x})\rightarrow
-i\hbar\,\frac{\delta}{\delta\phi(\mathbf{x})}$ which realizes the
commutation relation
$\left[\phi(\mathbf{x}),\pi(\mathbf{y})\right]=i\hbar\,\delta(\mathbf{x}-\mathbf{y})$.
The wave functional $\Psi[\phi]$ obeys the Schr\"{o}dinger
equation
    \beq
    i\hbar\frac{\partial}{\partial t}\Psi = \hat{H}\Psi
\,.
    \eeq
Since we will be considering perturbations in their ground states,
we take the wavefunctional to have a gaussian form
    \beq
    \Psi[\phi]=\mathcal{N} \exp\left[-\frac{1}{2}\int d^3x d^3y \,\phi(\mathbf{x})
    A(\mathbf{x},\mathbf{y},t)\phi(\mathbf{y})  \right]\,,
\label{Psi:ansatz}
    \eeq
with $\mathcal{N}$ the normalization factor. Due to the
homogeneity of the vacuum state, the correlator
in~(\ref{Psi:ansatz}) satisfies,
$A(\mathbf{x},\mathbf{y},t)=A(\mathbf{y},\mathbf{x},t)$. For this
state the Schr\"{o}dinger equation gives
    \bea \label{eom:N}
    i\hbar \, \partial_t\ln{\mathcal{N}}&=&\frac{\hbar^2}{2a^{3}}\int d^3x \,
    A(\mathbf{x},\mathbf{x}, t)  \\
    i\hbar\partial_tA(\mathbf{y},\mathbf{z},t)
  &=& \frac{\hbar^2}{a^3} \int d^3 x
               A(\mathbf{x},\mathbf{y},t)A(\mathbf{x},\mathbf{z},t)
               + a\left(\partial_\mathbf{y}^2-a^2m^2\right)
             \delta^{(3)}(\mathbf{y}-\mathbf{z})
\,.
\label{eom:A}
    \eea
Since the theory is noninteracting, it is convenient to
solve~(\ref{eom:A}) in momentum space, where it becomes local.
Indeed, upon writing
    \bea
    \phi(\mathbf{x})&=&\int \frac{d^3k}{(2\pi)^3} \, \phi_\mathbf{k}
    e^{i\mathbf{k}\cdot \mathbf{x}}\,,
\label{phi_k}
\\
    A(\mathbf{x},\mathbf{y},t)&=&\int \frac{d^3k}{(2\pi)^3} \, A(\mathbf{k},t)
    e^{i\mathbf{k}\cdot(\mathbf{x}-\mathbf{y})}\,,
    \eea
with $\phi_{-\mathbf{k}}=\phi_{\mathbf{k}}{}^\star$ and
$A(-\mathbf{k},t)=A(\mathbf{k},t)$, we get
    \beq
    i\hbar \, \partial_t
   A(\mathbf{k},t)=\frac{\hbar^2}{a^3}A^2(\mathbf{k},t)-a(k^2+a^2m^2)
\,.
\label{eom:kernel A}
    \eeq
The \emph{Heisenberg picture} mode functions
$\psi_\mathbf{k}(\eta)$ of a scalar field with the
action~(\ref{action}) satisfy
    \beq
    \partial_\eta^2 (a \psi_\mathbf{k}(\eta))
    +\left(k^2+\frac{\frac{1}{4}-\nu^2}{\eta^2}\right)
    (a\psi_{\mathbf{k}}(\eta))=0\,,
\label{eom:mode fns}
    \eeq
where $\eta$ is the conformal time, defined via $ad\eta=dt$, and
$\nu^2=\frac{9}{4}-\frac{m^2}{H^2}$. The Bunch-Davies vacuum
solution of~(\ref{eom:mode fns}) is of the form
    \beq
    a\psi_\mathbf{k}(\eta)
  = \sqrt{-\frac{\pi\eta}{4}}e^{i\frac{\pi}{2}\left(\nu + {1\over
    2}\right)}H_\nu^{(1)}(-k\eta)\,,
\label{mode fns:BD vacuum}
    \eeq
and the correlator $A(\mathbf{k})$~(\ref{eom:kernel A}) is related
to the mode functions~(\ref{mode fns:BD vacuum})
via~\cite{Guth:1985ya}
    \beq
    A(\mathbf{k},\eta)=\frac{1}{2 \hbar |\psi_\mathbf{k}(\eta)|^2}
            \left(1-i\, a^2\partial_\eta |\psi_\mathbf{k}(\eta)|^2
           \right)
\,.
\label{Avsphi}
    \eeq
At early times, in the regime $k^2\eta^2 \gg 1$ when the mode is
``deep inside the horizon'' and spacetime curvature is not
important, the solution~(\ref{mode fns:BD vacuum}) or
equivalently~(\ref{Psi:ansatz}) and (\ref{Avsphi}) reduces to that
of Minkowski vacuum. At late times, when $k^2\eta^2 \ll 1$, the
mode has ``exited the horizon'' and amplification (``particle
production") occurs. The wave function for each mode can be
written as\footnote{The reality of the field $\phi(\mathbf{x})$
implies that the modes $\phi_{-\mathbf{k}}$ and $\phi_{\mathbf{
k}}$ are related as $\phi_{-\mathbf{k}} =
\phi^\star_{\mathbf{k}}$, which necessitates that wave functions
be represented in terms of two-mode
states~\cite{Albrecht:1992kf}.}
    \beq
    \Psi(\phi_\mathbf{k},\eta) \propto\exp\left(-\frac{1}{2}\, \phi_\mathbf{k}A(\mathbf{k},\eta)
    \phi_\mathbf{k}^\star\right)\,,
\label{Psi:mode}
    \eeq
where the normalization factor can be easily obtained
from~(\ref{eom:N}) and by requiring $\int d\phi
d\phi^\star|\Psi|^2=1$. From now on we shall be ignoring such
factors. Note that in the Schr\"{o}dinger picture the time
evolution of the wave function is determined by the correlator
$A(\mathbf{k},\eta)$; $\phi_\mathbf{k}$ is \emph{time independent}
and should not be confused with the Heisenberg picture mode
$\psi_\mathbf{k}(\eta)$~(\ref{mode fns:BD vacuum}).

The quantum mechanics of cosmological perturbations has been
studied in~\cite{Guth:1985ya,Albrecht:1992kf}. The system evolves
into a kind of state known as a squeezed state, which exhibits a
high degree of WKB classicality
    \beq\label{WKB}
    \hat{p}\,\Psi \simeq \partial_qS(q)\Psi\,,
    \eeq
where $\hat{q}$ and $\hat{p}$ are the configuration variable and
its conjugate momentum, while $S$ is the exponent of the wave
function, practically the classical action. Thus, the amplitude of
each mode and its conjugate momentum are related to a very high
degree of accuracy via the corresponding classical relation. In
cosmology, this state of affairs is interpreted as equivalent to a
statistical mixture of mode amplitude eigenstates, where the
probability that any of these states has been realized in our
universe is given by $|\Psi(\phi_\mathbf{k},\eta)|^2$. This would
correspond to a density matrix which is diagonal in the field
amplitude basis
    \beq
    \hat{\rho}_\mathbf{k}=\sum\limits_{\phi_\mathbf{k}}
                              |\Psi(\phi_\mathbf{k},\eta)|^2
    \,|\phi_\mathbf{k}\rangle\langle\phi_\mathbf{k}|.
\label{rho:mixed description}
    \eeq
Such a stochastic mixture is a very good approximation when
calculating correlators such as $\langle\phi_\mathbf{k}^2\rangle$
\cite{Polarski:1995jg}, which are then related to the stochastic
properties of cosmological perturbations. However, the wave
function~(\ref{Psi:mode}) does not directly lead to the
interpretation suggested by~(\ref{rho:mixed description}).
Explicitly calculating the density matrix from~(\ref{Psi:mode}),
one finds
    \beq
    \langle\phi_\mathbf{k}|\hat{\rho}_\mathbf{k}|\bar{\phi}_\mathbf{k}\rangle
    \equiv  \rho(\phi_\mathbf{k},\bar{\phi}_\mathbf{k})=
    \Psi(\phi_\mathbf{k},\eta) \Psi^\star(\bar{\phi}_\mathbf{k},\eta)  \\
     \propto\exp \big(-u_\mathbf{k}\Re[{A}]u_\mathbf{k}^\star \!- \frac{1}{4}\Delta_\mathbf{k}
    \Re[{A}]
    \Delta_\mathbf{k}^\star -i\Im[{A}]\Re[u_\mathbf{k}\Delta_\mathbf{k}^\star]\big)  \nonumber,
    \eeq
where we have defined $u=(\phi +\bar{\phi})/2$ and $\Delta=\phi -
\bar{\phi}\,$. The above expression does not vanish for $\Delta
\neq 0$, making the off-diagonal terms of the density matrix
non-zero. Thus, strictly speaking, the description~(\ref{rho:mixed
description}) is not valid and one cannot say that any particular
mode amplitude has been realized with a certain probability. The
modes of cosmological perturbations exist in coherent
superpositions with different mode amplitudes. In order to
diagonalize the density matrix, some process of decoherence must
take place. As we show below, the existence of fields other than
the inflaton during inflation can decohere the density matrix of
cosmological perturbations even if no direct coupling between the
different fields is assumed. The relevant coupling is ensured by
the existence of scalar metric perturbations.

\section{Isocurvature perturbations and decoherence}

\subsection{Adiabatic and isocurvature perturbations}
The existence of isocurvature modes is ubiquitous when more than one
scalar fields are relevant during inflation. We will therefore consider
a two-field model of inflation with a potential $V(\varphi,\chi)$. For
cosmological perturbations for the metric and the scalar fields
    \bea
    \phi(\mathbf{x},t)=\phi(t)+\delta\phi(\mathbf{x},t),\quad
    \chi(\mathbf{x},t)=\chi(t)+\delta\chi(\mathbf{x},t)\,,\\
    ds^2=-(1+2\Phi(\mathbf{x},t))dt^2 +
    a^2(t)(1-2\Phi(\mathbf{x},t))d\vc{x}^2\,,
    \nonumber
    \eea
the Lagrangian describing the dynamics of gauge invariant
perturbations can be written
as \cite{vanTent}
    \beq
    L=\int \!\!d^3x \,\,\frac{1}{2}\partial_\eta\mathbf{q}\partial_\eta\mathbf{q} -
    \frac{1}{2}\mathbf{q}\left(-\nabla^2+(aH)^2\mathbf{\Omega}\right)\mathbf{q}\,,
    \eeq
where
    \beq
    \mathbf{q}=
    a(\delta\varphi+\frac{\dot{\varphi}}{H}\Phi)\mathbf{\hat e}_\varphi
    + a(\delta\chi+\frac{\dot{\chi}}{H}\Phi)\mathbf{\hat e}_\chi
    \equiv q_\varphi \mathbf{\hat e}_\varphi + q_\chi \mathbf{\hat
    e}_\chi\,,
    \eeq
with $\mathbf{\hat e}_\varphi$ and $\mathbf{\hat e}_\chi$ the unit
basis vectors in the $\varphi$ - $\chi$ field coordinates. The matrix
$\mathbf{\Omega}$ is given by
    \beq
    \mathbf{\Omega} = \frac{\partial_a\partial_bV}{H^2}\mathbf{
    e}_a\otimes \mathbf{e}_b - (2-\epsilon)\mathbf{1} - 2 \epsilon\left( \left(3+\epsilon\right)\,\mathbf{e}_1\otimes
    \mathbf{e}_1+2\eta^\|\mathbf{e}_1\otimes\mathbf{e}_1
    +\eta^\perp \left(\mathbf{e}_1\otimes\mathbf{e}_2 +\mathbf{e}_2\otimes\mathbf{e}_1\right) \right)\,,
    \eeq
where $\epsilon=-\dot{H}/H^2$ and
    \beq
    \eta^\| \equiv  -3-\frac{\partial\mathbf{V}\cdot
    \mathbf{e}_1}{H|\dot{\phi}|}\,,\quad \eta^\perp \equiv  -\frac{\partial\mathbf{V}\cdot
    \mathbf{e}_2}{H|\dot{\phi}|}\,,
    \eeq
are two slow roll parameters related to the acceleration parallel and
perpendicular to the field velocity \cite{vanTent}. The subscripts $\{a,b\}$
take the values $\{\varphi,\chi\}$ and we have used the unit vectors
$\mathbf{\hat e}_1$ and $\mathbf{\hat e}_2$ to be defined shortly. Note
that even if scalar fields are decoupled and the potential term is
diagonal, the $q_a$ fields are still coupled due to the inclusion of
the metric perturbation $\Phi$. Thus, perturbations are always coupled
gravitationally. As we discuss below this coupling can suffice to
decohere the perturbations if one degree of freedom is traced out.

In order to proceed, it is convenient to define two new basis
vectors, tangential and normal to the field trajectory
\cite{vanTent, Gordon}:
    \bea
    \mathbf{e}_1 &=& \cos\theta \, \mathbf{e}_\varphi
    + \sin\theta \, \mathbf{e}_\chi \equiv \frac{\dot{\varphi}}{\sqrt{ \dot{\varphi}^2+\dot{\chi}^2 }}
    \, \mathbf{e}_\varphi
    + \frac{\dot{\chi}}{\sqrt{ \dot{\varphi}^2+\dot{\chi}^2}}\,\mathbf{
    e}_\chi\\
    \mathbf{e}_2 &=& -\sin\theta\, \mathbf{e}_\varphi +
    \cos\theta\,\mathbf{e}_\chi \,,
    \eea
and express the Lagrangian in terms of components in this basis -
we have
    \beq
    L=\int\!\!d^3x\,\,\frac{1}{2}\left(\partial_\eta\s{q}
    +\s{Z}\s{q}\right)^\s{T}\left(\partial_\eta\s{q}+\s{Z}\s{q}\right)
    -\frac{1}{2}\s{q}^\s{T}\left(-\nabla^2+(aH)^2\s{\Omega}\right)\s{q}\,,
    \eeq
where $(m,n=1,2)$
    \beq
    \s{q}=\begin{pmatrix}q_1 \\ q_2 \end{pmatrix}\,, \quad
    \s{q}_m=\mathbf{e}_m \cdot \mathbf{q}\,,
    \eeq
    \beq
    \s{\Omega}_{mn}=\mathbf{e}_m\mathbf{\Omega}\mathbf{e}_n
    \eeq
is the matrix $\mathbf{\Omega}$ expressed in the $\{\mathbf{\hat
e}_1 \,, \mathbf{\hat e}_2\}$ basis and
    \beq
    \s{Z}=\partial_\eta\theta\begin{pmatrix}0&-1\\1&0\end{pmatrix}\,.
    \eeq
Then, using the conjugate momentum
$\pi(\vc{x})=\frac{\partial{\mathcal{L}}}{\partial (\partial_\eta q(\vc{x}))}$ the
Hamiltonian can be written as
    \beq
    H=\int\!\!d^3x\,\,\frac{1}{2}\s{\pi}^\s{T}\s{\pi} - \frac{1}{2}\s{\pi}^\s{T}\s{Z}\s{q}
    - \frac{1}{2}\s{q}^\s{T}\s{Z}^\s{T}\s{\pi} +
    \frac{1}{2}\s{q}^\s{T}\left(-\nabla^2+(aH)^2\s{\Omega}\right)\s{q}\,.
    \eeq
The quantum Hamiltonian is obtained via the replacement
$\s{\pi}(\vc{x}) \rightarrow -i\hbar \frac{\delta}{\delta
q(\vc{x})}$ and the wavefunction satisfies the Schr\"{o}dinger
equation
    \beq
    i\hbar\Psi=\hat{H}\Psi\,,
    \eeq
which, along with the vacuum ansatz
    \beq\label{wavefn}
    \Psi=N\exp{\left(-\frac{1}{2}\int\!\!d^3xd^3y
    \,\,\s{q}^\s{T}(\vc{x})\s{B}(\vc{x}-\vc{y})\s{q}(\vc{y})\right)}\,,
    \eeq
leads to
    \bea\label{schrodinger-matrix}
    i\hbar\partial_\eta\s{B}(\vc{x}-\vc{y})&=&\hbar^2\left(\int d^3z\,\s{B}(\vc{x}-\vc{z})\s{B}(\vc{z}-\vc{y})\right)
    + i\hbar
    \left[\s{B}(\vc{x}-\vc{y}),\,\s{Z}\right] +
    \left(\nabla_\vc{x}^2-(aH)^2\s{\Omega}\right)\delta^{(3)}(\vc{x}-\vc{y})\,,\\
    i\hbar\partial_\eta\ln{N}&=&\frac{\hbar}{2}\int d^3x
    \,{\s{B}(0)}\,.
    \eea
The solution to these equations provides the full quantum description
of cosmological perturbations in the two-field system.

\subsection{The reduced density matrix}

The process of decoherence relies on the existence of degrees of
freedom which couple to - get entangled with - the system of
interest but which are not directly accessible. It is assumed that
they cannot be measured along with the system and thus provide an
uncontrolable environment. Tracing them out leads to the
diagonalization of the density matrix for the system of interest
and thus explains the appearance of a classical world where
different alternatives - like Schr\"{o}dinger's cat being dead and
alive - do not interfere and are perceived as distinct
\cite{Zurek, Joos}. In Cosmology, the role of the cat is played by
the long wavelength fluctuations which lead to structure
formation.

In what follows we suggest that decoherence can occur for
cosmological perturbations if the isocurvature component $q_2$ is
considered to be the unobservable environment and is traced out.
In cosmological considerations the two degrees of freedom in
$\mathbf{q}$ describing perturbations are not directly observable.
A more tangible remnant of inflation is the curvature perturbation
which persists in later epochs of cosmic evolution and is the
reason for structure formation. More precisely, it is given by a
scalar quantity $\mathcal{R}$ which parameterizes the spatial
curvature of comoving time slices (see e.g. \cite{liddle-lyth}).
During inflation it is related to $\s{q}_1$ by
    \beq
    \hat{\mathcal{R}}=\frac{H}{\sqrt{\dot{\varphi}^2+\dot{\chi}^2}}\,\frac{\hat{\s{q}}_1}{a}\,,
    \eeq
and for single field inflation it is constant on long
weavelengths. The presence of an isocurvature perturbation makes
$\mathcal{R}$ evolve in time \cite{Gordon, vanTent}
    \beq\label{dotR}
    \frac{d}{dN}\hat{\mathcal{R}}=2\frac{ {\partial V} \cdot\mathbf{e}_2}
    {\dot{\varphi}^2+\dot{\chi}^2}\,\frac{\hat{\s{q}}_2}{a}\,.
    \eeq
The hat acts as a reminder that we are dealing with quantum operators.

Any process which couples to $\hat{\mathcal{R}}$ during inflation
is effectively measuring $\hat{q}_1$, the adiabatic component of
the perturbations. A measurement of $\hat{q}_2$ via gravitational
interactions alone would entail a coupling to
$\frac{d}{dN}\hat{\mathcal{R}}$ during inflation as can be seen
from (\ref{dotR}). Thus, the gravitational perturbation
$\hat{\mathcal{R}}$, which does not have its own independent
dynamics, acts as an apparatus which``measures" - gets entangled
with - $\hat{q}_1$. Let us now assume that after inflation both
$\phi$ and $\chi$ decay into relativistic particles, generating no
entropy perturbation and making $\hat{\mathcal{R}}$ the only
remnant from the inflationary epoch.\footnote{This reasoning is
not valid if a gravitational potential is generated after
inflation via a curvaton type
mechanism~\cite{Lyth:2001nq,Enqvist:2001zp}.} In this case,
information about $\frac{d}{dN}\hat{\mathcal{R}}$ during inflation
is lost and all post-inflationary perturbations can be obtained
from $\hat{\mathcal{R}}$ at the end of inflation, which is
conserved on super-Hubble scales after $\hat{\s{q}}_2$ decays.
Once perturbations reenter the horizon at late times,
$\frac{d}{dN}\hat{\mathcal{R}}$ becomes different from zero again.
This late time behaviour reflects late time dynamics of matter
fields on small scales and has nothing to do with the inflationary
$\frac{d}{dN}\hat{\mathcal{R}}$. Consequently, as information
about $\hat{q}_2$ is lost, we can consider it to be an
inaccessible environment which cannot be directly observed and can
therefore be traced out of the density matrix.

Under the above assumption, the density matrix relevant for
the adiabatic perturbation alone is the one obtained by tracing out
$q_2$
     \beq
    \tilde{\rho}(q_1,  \bar{q}_1)=\int
    dq_2 dq_2^\star\Psi(q_1,
    q_2)\Psi^\star(\bar{q}_1,q_2)\,,
    \eeq
where the wavefunction $\Psi$ is given by (\ref{wavefn}). Expressing the result in terms of the variables $u=(q_1
+\bar{q}_1)/2$ and $\Delta=q_1 - \bar{q}_1$, we find
    \beq
    \tilde{\rho}(u,\Delta)
   =\exp\left[-\frac{1}{2}(u,\Delta)\mathbf{\s{C}}\left(\begin{matrix}u^\star
 \\
   \Delta^\star\end{matrix}\right)\right]\,,
\label{density matrix:reduced}
    \eeq
with
    \beq
    \s{C}\!\equiv\!\left(\begin{matrix}
    2\left(\Re[\s{B}_{11}]-\frac{\Re[\s{B}_{12}]^2}{\Re[\s{B}_{22}]}\right) &
 \!{i}\left(\Im[\s{B}_{11}]\!-\!\frac{\Re[\s{B}_{12}]\Im[\s{B}_{12}]}{2\Re[\s{B}_{22}]}\right)\!
    \\
 \!{i}\left(\Im[\s{B}_{11}]\!-\!\frac{\Re[\s{B}_{12}]\Im[\s{B}_{12}]}{2\Re[\s{B}_{22}]}\right)\!
    & \frac{1}{2}\left(\Re[\s{B}_{11}]+\frac{\Im[\s{B}_{12}]^2}{\Re[\s{B}_{22}]}\right)
    \end{matrix}\right)\,.
\label{Dmatrix}
    \eeq

As we remarked above, decoherence would imply that the $\Delta$
components are suppressed when $|\Delta| \neq 0$. This will happen if
    \beq\label{condition}
    \s{C}_{11}/\s{C}_{22}\ll 1\,,
    \eeq
since in this case the relative importance of the $\Delta$ over the $u$
terms will diminish, making quantum correlations unimportant. Thus, we
expect that decoherence will become effective if
    \beq\label{deco condition}
    \Im[\s{B}_{12}] > 2 \sqrt{\Re[\s{B}_{11}]\Re[\s{B}_{22}]}\,.
    \eeq
In inflationary models with two scalar
fields, $\Omega_{12}$ in
(\ref{schrodinger-matrix}) is in general non zero, $\Omega_{12}\neq 0$ and thus $\s{B}_{12}\neq 0$ (see also (\ref{schrodinger-3}) and
(\ref{omegatheta})). In particular, for inflationary perturbations and on superhorizon scales, the imaginary part $\Im{[B_{12}]}$ strongly dominates over the real parts $\Re{[B_{11}]}$ and
$\Re{[B_{22}]}$ and thus, when isocurvature modes are inaccessible, we expect that decoherence will occur generically.\footnote{For the behavior of real and imaginary parts relating to
inflationary perturbations see \cite{Polarski:1995jg} and the next
section} We check this assertion explicitly in the next section for a simple two field model with
two massive free fields. We find that this expectation is
indeed correct for a relatively wide range of parameter values of the
model.

\section{Decoherence in the model  $V(\varphi,\chi)=\frac{1}{2}m^2\varphi^2+\frac{1}{2}\mu^2\chi^2$}

In this section we illustrate the role of isocurvature perturbations
for decoherence in an inflationary model with
$V(\varphi,\chi)=\frac{1}{2}m^2\varphi^2+\frac{1}{2}\mu^2\chi^2$, a
model which has been studied extensively in \cite{polarski2} (see also
\cite{vanTent}). We first present exact numerical results for two
cases: a) $\mu=1.05 m$ and b) $\mu=10m$. We then derive an analytic
result for the first case. We note that the first case is practically
the same as single-field inflation as the trajectory in field space is
practically straight. In the second case, the heavier field drops
out of slow roll and start oscillating approximately 40 e-folds before
inflation ends. However, as cosmologically observable scales exit the horizon
while both fields are still in slow roll, the spectral index does not
deviate significantly from unity and thus the model is observationally
viable.

The wavefunction obtained from (\ref{schrodinger-matrix}) explicitly
reads:
    \bea\label{schrodinger-1}
    i\hbar \partial_\eta
    \s{B}_{11}&=&\hbar^2\left(\s{B}_{11}^2+\s{B}_{12}^2\right)+2i\hbar\partial_\eta\theta
    \s{B}_{12} - \left(k^2 +
    (aH)^2\s{\Omega}_{11}\right)\,,\\ \label{schrodinger-2}
    i\hbar \partial_\eta \s{B}_{22}&=&\hbar^2\left(\s{B}_{22}^2+\s{B}_{12}^2\right) - 2i\hbar\partial_\eta\theta
    \s{B}_{12} - \left(k^2 + (aH)^2\s{\Omega}_{22}\right)\,,\\
    \label{schrodinger-3}
    i\hbar \partial_\eta \s{B}_{12}&=&\hbar^2\s{B}_{12}\left(\s{B}_{11}+\s{B}_{22}\right)
    + i\hbar\left(\s{B}_{22}-\s{B}_{11}\right)\partial_\eta\theta
    - (aH)^2\s{\Omega}_{12}\,,
    \eea
where
    \bea
    \s{\Omega}_{11} &=&-2+
    \frac{1}{2H^2}\left(\mu^2+m^2 - \left(\mu^2-m^2\right)\cos 2\theta\right)
    -5\epsilon-2\epsilon\left(\epsilon+2\eta^\|\right)\,,\\
    \s{\Omega}_{22} &=& -2 +
    \frac{1}{2H^2}\left(\mu^2-m^2 + \left(\mu^2+m^2\right)\cos 2\theta\right)+ \epsilon \,,\\
    \s{\Omega}_{12} &=& \s{\Omega}_{21} =
    \frac{\mu^2-m^2}{2H^2}\sin{2\theta}-2\epsilon\eta^\perp 
    \label{omegatheta}
    \eea
These equations are solved numerically with initial condition
$\s{B}_{11}=\s{B}_{22}\simeq{k}/{\hbar}$, $B_{12}=0$. Figure 1 shows
the evolution of
    \beq\label{deco ratio}
    s_\vc{k}\equiv{1\over 2}\ln\frac{4\s{C}_{22}}{\s{C}_{11}}={1\over 2}\ln\left[\frac{\Re[B_{11}]\Re[B_{11}]+\Im[B_{12}]^2}{\Re[B_{11}]\Re[B_{11}]-\Re[B_{12}]^2}\right],
    \eeq
which controls the amount of decoherence, for the two mass ratios
$\mu/m=1.05$ (blue, solid line) and $\mu/m=10$ (black, solid line). This ratio is also the entropy per mode, see section V. Horizon exit occurs at 5 and 3 efolds respectively for the chosen mode and
inflation finishes at 60 efolds. We see that $s_{\vc{k}}$ becomes larger than
unity a few efolds after horizon exit and continues to grow
linearly with the number of e-fold causing the density matrix to decohere. For $\mu/m=1.05$
this growth continues until the end of inflation. For the larger mass
ratio it stops once the heavier field drops out of slow roll and starts
oscillating, but $s$ retains its average value until the end of
inflation (it becomes zero for very short intervals and is quickly
restored). Thus, in both cases decoherence is established a few efolds
after horizon crossing and is retained to the end of inflation.

The period of oscillations in figure 1 represents the Poincar\'e recurrence
time for our system. The recurrence time is short because for each mode
we have only two degrees of freedom. In models with several scalar fields
however, we expect that the Poincar\'e recurrence time grows exponentially.
Hence the brief recoveries of coherence we see in figure 1 should be a specific
feature of the two scalar field model under consideration.

\begin{figure}[t]
    \includegraphics[width=11cm]{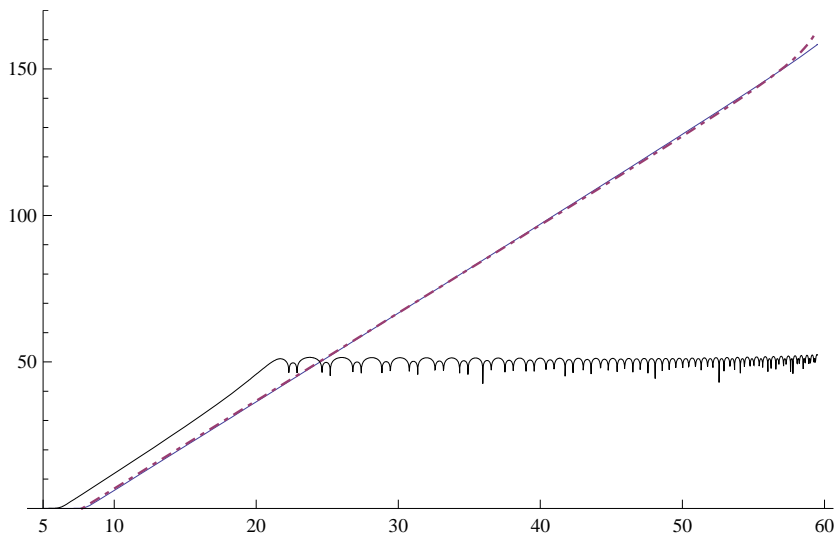}
    \caption{The entanglement entropy per mode $s_\vc{k}$ plotted for the cases $\mu=1.05m$ (numeric: solid, blue line; analytic: dashed-dotted, red line) and $\mu=10m$ (numeric: solid, black line). For the first case, the analytic and numerical results are hardly distinguishable. Horizon exit happens at 5 and 3 e-folds respectively for the two cases and a few e-folds later the density matrix has decohered. For similar masses the entropy grows linearly with the number of e-folds until the end of inflation. For dissimilar masses, the entropy stops growing when the heavy field starts oscillating and remains constant on average, becoming zero for negligible time intervals.}
\end{figure}

The case of almost equal masses can also be treated analytically where
both fields exhibit slow-roll:
    \beq
    \frac{d\phi}{dN} \simeq  \frac{\partial_\varphi V}{3H^2} \,,
 \quad \frac{d\chi}{dN} \simeq \frac{\partial_\chi V}{3H^2}\,,\quad
 H^2 \simeq \frac{1}{3M_p^2}V
 \label{dot phi chi}
    \eeq
with $dN=-H dt=-aHd\eta$, $M_p^2=(8\pi G)^{-1}$, $H=\dot{a}/a$ the
(almost constant) inflationary expansion rate of the universe. Denoting
the number of e-folds N as
    \beq
    a(N)=a_\star e^{N_\star-N}\,,
    \eeq
we have that
    \bea
    \varphi&=&2M_p\sqrt{N}\cos\omega\\
    \chi &=&2M_p\sqrt{N}\sin\omega\,.
    \eea
As N decreases inflation proceeds, ending when $N \simeq 1$. The anglular variable $\omega$ evolves according to
    \beq
    \frac{d\omega}{dN} = \frac{\mu^2-m^2}{6H^2}\sin{2\omega}\,.
    \eeq
Slow roll implies that
    \beq
    \tan{\theta}=\frac{\mu^2}{m^2}\tan{\omega} \,,
    \eeq
    which gives
    \beq
    \frac{d\theta}{dN} =
    \frac{\mu^2-m^2}{6H^2}\sin{2\theta}\,.
    \eeq
The above expression can be easily integrated (see
\cite{polarski2} for details). The expansion rate $H$ is then
given by
    \beq\label{H}
    H^2=\frac{1}{3}N\left(\mu^2+m^2-\left(\mu^2-m^2\right)\cos{2\omega}\right)\,.
    \eeq
Furthermore, we are going to need an expression for the slow roll parameter
    \beq\label{epsilon}
    \epsilon=-\dot{H}/H^2=\frac{1}{2M_p^2}\left[\left(d\varphi/dN\right)^2+\left(d\chi/dN\right)^2\right]\,.
    \eeq
We have
    \beq\label{epsilon2}
    \epsilon = \frac{1}{N} \frac{m^4+\mu^4
    -\left(\mu^4-m^4\right)\cos{2\omega}}{\left[\mu^2+m^2-\left(\mu^2-m^2\right)\cos{2\omega}\right]^2}\,.
    \eeq

Let us first focus on subhorizon scales, where the
$\partial_\eta\theta$ terms are not important; we have
    \bea\label{schrodinger-short}
    i\hbar \partial_\eta
    \s{B}_{11}&=&\hbar^2\left(\s{B}_{11}^2+\s{B}_{12}^2\right) - k^2 \,,\\
    i\hbar \partial_\eta \s{B}_{12}&=&\hbar^2 2 \s{B}_{12}\s{B}_{11}
    - (aH)^2\s{\Omega}_{12}\,,
    \eea
with solution $\s{B}_{11}=\s{B}_{22}\simeq{k}/{\hbar}$, $B_{12}=0$,
which is the standard Minkowski vacuum for two decoupled degrees of
freedom.

On superhorizon scales, an analytic solution to (\ref{schrodinger-1}) -
(\ref{schrodinger-3}) can be obtained perturbatively if
$\partial_\eta\theta$ is treated as a small parameter. This will be
true if
    \beq\label{mass condition}
    \frac{\mu^2-m^2}{m^2}\ll 1.
    \eeq
To $0^{th}$ order in $\partial_\eta\theta$ we have
    \bea\label{schrodinger-long1}
    i\hbar \partial_\eta
    \s{B}_{11}&\simeq&\hbar^2\left(\s{B}_{11}^2+\s{B}_{12}^2\right) - \left(k^2 +
    (aH)^2\s{\Omega}_{11}\right)\,,\\ \label{schrodinger-long2}
    i\hbar \partial_\eta \s{B}_{22}&\simeq&\hbar^2\left(\s{B}_{22}^2+\s{B}_{12}^2\right)
     - \left(k^2 + (aH)^2\s{\Omega}_{22}\right)\,,\\
    \label{schrodinger-long3}
    i\hbar \partial_\eta
    \s{B}_{12}&\simeq&\hbar^2\s{B}_{12}\left(\s{B}_{11}+\s{B}_{22}\right)\,.
    \eea
Thus, $\s{B}_{12}=0$ and $B_{11}$, $\s{B}_{12}$ can be obtained
from the single field results (see (\ref{Avsphi})) where $B(\vc{k},\eta)\equiv
A(\vc{k},\eta)/a^2$ and in the long wavelength limit
    \beq\label{longwave}
    \lim \limits_{k\ll aH} q_{ k} \simeq \frac{e^{i{\pi\over 2}\left(\nu-{1\over
    2}\right)}}{\sqrt{2k\pi}}\Gamma(\nu)\left(-\frac{k\eta}{2}\right)^{-\nu+{1\over
    2}}\,.
    \eeq
Using these, we find to leading order for the correlator
    \bea
    \s{B}_{11}=\frac{aH}{\hbar} \left(  \frac{2\pi}{\Gamma(\nu_1)^2}\left(\frac{k}{2aH} \right)^{2\nu_1}
    \!\!+ i\, \left( \frac{3}{2}-\nu_1 \right)  \right)\,,
    \label{A:expansion1} \\
    \s{B}_{22}=\frac{aH}{\hbar} \left(  \frac{2\pi}{\Gamma(\nu_2)^2}\left(\frac{k}{2aH} \right)^{2\nu_2}
    \!\!+ i\, \left( \frac{3}{2}-\nu_2 \right)  \right)\,,
    \label{A:expansion2}
    \eea
where
    \bea \label{nu1}
    \nu_1 &\simeq& {3 \over 2} +{5\over 3} \epsilon - \frac{1}{6H^2}\left(\mu^2+m^2
    - \left(\mu^2-m^2\right)\cos 2\theta_0\right)\\
    \nu_2 &\simeq& {3\over 2} - {1\over 3}\epsilon - \frac{1}{6H^2}\left(\mu^2-m^2
    + \left(\mu^2+m^2\right)\cos 2\theta_0\right)\,,
    \eea
with $\tan{\theta_0} = \frac{\mu^2}{m^2}\tan{\omega_0}$. Then, to
first order in $\partial_\eta\theta$ we have
    \beq
    i\hbar\partial_\eta \s{B}_{12} \simeq \hbar^2\s{B}_{12}\left(\s{B}_{11}+\s{B}_{22}\right)+
    \left[i\hbar\left(\s{B}_{22}-\s{B}_{11}\right)+3aH\right]\partial_\eta\theta\,.
    \eeq
Soon after horizon crossing, $\s{B}_{11}$ and $\s{B}_{22}$ become imaginary and $\s{B}_{12}$ is given by
    \beq\label{B12}
    \s{B}_{12} = - \frac{i}{\hbar} \int\limits^N_{N_\star}  e^{\int^N_{\bar{N}}\!\!\alpha d{N'}}\, (3-\beta)\,
    aH \,\frac{\mu^2-m^2}{6H^2}\sin2\theta\,d\bar{N}\,,
    \eeq
where we have taken $N_\star$ to be a time when the real parts of
$\s{B}_{11}$ and $\s{B}_{22}$ have become negligible, and
    \bea
    \alpha  &=&  {4\over 3}\epsilon - \frac{\mu^2+m^2\cos2\theta_0}{3H^2} = \frac{\gamma}{N}\,,\\
    \beta   &=&  2\epsilon + \frac{\mu^2\cos2\theta_0-m^2}{3H^2}\cos 2\theta_0 =\frac{\delta}{N}
    \,.
    \eea
Using (\ref{H}) and (\ref{epsilon2}), and neglecting the dependence
of $\theta$ on $N$ in (\ref{B12}) we find
    \beq\label{B12approx}
    \s{B}_{12} (N) \simeq \frac{i}{\hbar}\frac{\mu^2-m^2}{2\sqrt{3}}
    \frac{\sin{2\theta_0}}{\sqrt{\mu^2+m^2-\left(\mu^2-m^2\right)\cos{2\omega_0}}}
    \,a(N) I(N)\,,
    \eeq
where
    \bea
    I(N)&\equiv& N^{\gamma}\int\limits^{\,\,\,N_\star}_{N}
    \left(3-\frac{\delta}{s}\right)\frac{e^{N-s}}{s^{\gamma+\frac{1}{2}}}\,\,ds  \nonumber \\
    &=& N^\gamma e^{N}\left(\delta\,\Gamma(-{1\over2}-\gamma,N_\star)-3\,\Gamma({1\over2}-\gamma,N_\star)
    -\delta\,\Gamma(-{1\over 2}-\gamma,N)+3\,\Gamma({1\over 2}-\gamma,N)\right)\,.\label{integral}
    \eea
As seen in figure 1, the above formulae reproduce the relevant numerical results with high accuracy.


\section{Entanglement Entropy}

 \begin{figure}[t]
    \includegraphics[width=7cm]{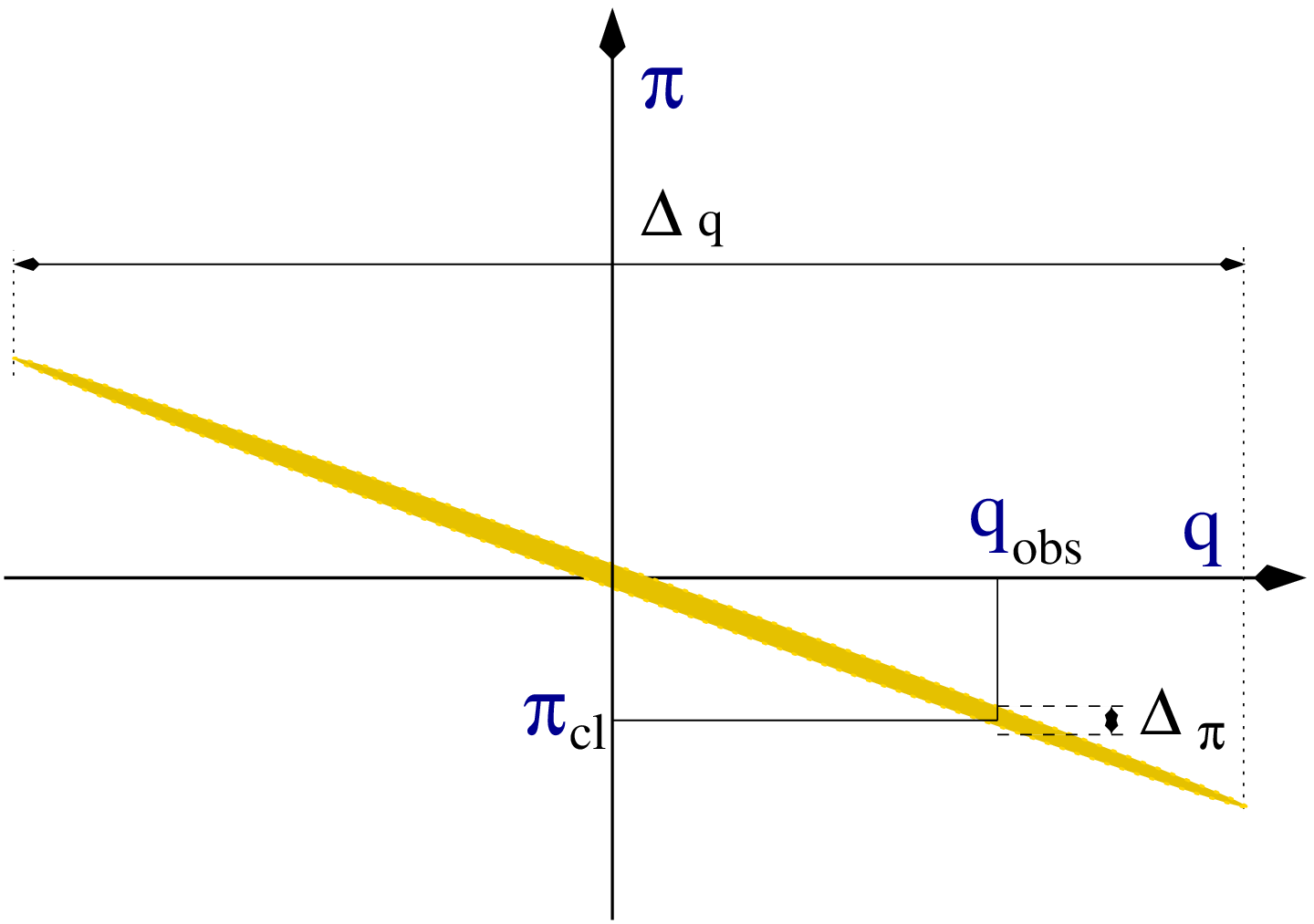}
    \includegraphics[width=7cm]{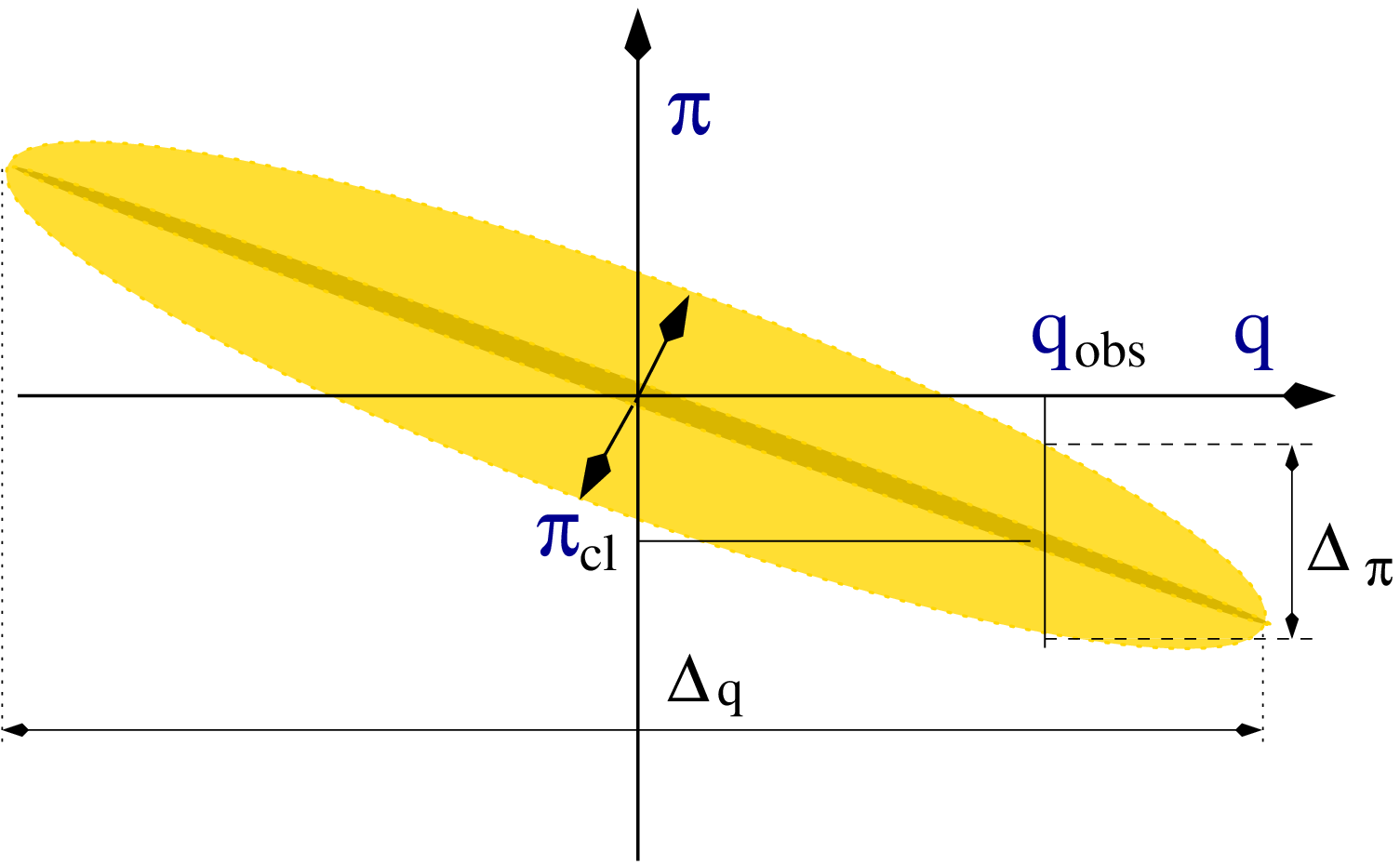}
    \caption{{\it Left}: An illustration of the Wigner function before decoherence for a highly
    squeezed state describing long wavelength cosmological quantum fluctuations.
    The field amplitude (horizontal axis) is highly correlated with the field momentum (vertical axis).
    The area of the ellipse corresponds to minimum uncertainty and the entropy is zero. {\it Right}:
    The Wigner function after decoherence. The ellipse has increased its area leading to a
    non-vanishing value for the entropy. Note that the correlation
between field amplitude and momentum has become less sharp but
since $\Delta_{\pi(deco)}/\pi_{class}\ll 1$ even after
decoherence, the prediction for acoustic oscillations in the CMB
is not spoiled.}
    \label{fig}
    \end{figure}

We now discuss how decoherence can be quantified by the amount of
entanglement entropy for cosmological perturbations which is generated
after tracing out one of the degrees of freedom. The analysis
in~\cite{Brandenberger:1992jh, Prokopec:1992ia} implies that the
entropy of a gaussian quantum state can be expressed in terms of the
product of the field amplitude and momentum variances ($\Delta_q^2$,
$\Delta_\pi^2$) as
    \begin{eqnarray}
    S &=& {\rm tr}\,{\rm \ln}
       \left[2\Delta_\s{q}\Delta_\s{\pi}/\hbar\right]
    \,,
    \label{entropy:general}
    \end{eqnarray}
with $S=0$ for any minimum uncertainty state such as the state of
cosmological perturbations. The variance of the momentum $\Delta_\pi$
is calculated from the Wigner function $W(q,\pi)$ corresponding to the
density matrix:
    \beq
    W(q,\pi)=\int d\pi d\pi^\star \,\rho\,
    e^{-i\Delta\pi^\star-i\Delta^\star\pi}\,,
    \eeq
which for a gaussian state reads
    \beq
    W(q,\pi) \propto \exp \big( -\s{q}\frac{1}{\Delta^2_\s{q}}\s{q}^\star
     - \left(\s{\pi}-\s{\pi}_{cl}\right)\frac{1}{\Delta^2_\s{\pi}}\left(\s{\pi}-\s{\pi}_{cl}\right)^\star \big)\,.
    \eeq
The classical momentum associated with a value for $\s{q}_1$ is
    \beq\label{class-mom}
    \s{\pi}_{1cl}=- \s{q}_1\s{C}_{12}\hbar\,.
    \eeq
Note that the existence of a classical momentum, corresponding to a
Wigner ellipse rotated in phase space (figure 2), does not alter the
variances of $\s{q}$ and $\s{\pi}$ and equation (\ref{entropy:general})
always holds as can be checked explicitly.

Before decoherence, $\Delta_{q_1}^2 = 1/\Re[\s{B}_{11}]$
and $\Delta_{\pi_1}^2 = \hbar^2\Re[\s{B}_{11}]/4$ such that the
state~(\ref{Psi:mode}) corresponds to the minimum uncertainty
state with vanishing entropy.
Calculating the entanglement entropy corresponding to the reduced density
matrix~(\ref{density matrix:reduced}),
we get for the decohered adiabatic perturbation
    \beq\label{entropy per mode}
    S =   V\int \frac{d^3k}{(2\pi)^3}\, s_{\mathbf{k}}=
    V\int
    \frac{d^3k}{(2\pi)^3}\,\frac12\ln\left[\frac{4\s{C}_{22}}{\s{C}_{11}}\right]\,.
    \eeq
Thus, the entropy is characterized by the same quantity (\ref{deco
ratio}) which controls decoherence and figure 1 also shows the evolution
of the entropy for the cases of the two mass ratios.

Equation~(\ref{entropy per mode}) establishes a link between the
entropy of cosmological perturbations and decoherence; the
entropy~(\ref{entropy per mode}) corresponds to the entanglement
entropy which equals (minus) the information stored in
correlations between the adiabatic and isocurvature perturbations.
Therefore, it conforms with the standard definition of the entropy
of quantum systems: tracing out the isocurvature perturbation (the
unobservable ``environment'') generates the entropy of the
adiabatic perturbation (the system).

The effect of decoherence on the Wigner function is illustrated in
figure 1. As mentioned above, prior to decoherence the state is
described by the squeezed ellipse, the area of which is dictated by the
minimum uncertainty. After decoherence, the variance $\Delta_{\s{q}_1}$
remains to a good accuracy unchanged but the variance in the momentum has increased to $\Delta^2_{\s{\pi_1}(dec)} = 2\s{C}_{22}$. This is opposite to the standard treatment of inflationary perturbations where the extremely squeezed nature of the state is taken to imply an exact classical correlation between the mode amplitude, drawn from a gaussian distribution, and the momentum. Decoherence implies that the momentum should also be drawn from a distribution with a dispersion larger than that dictated by the minimum quantum uncertainty.

The existence of acoustic peaks in the CMB requires that the field amplitude and the corresponding momentum exhibit a high degree of classical correlation as was already pointed out in \cite{kiefpolstar}. Assuming $\s{q_1}\sim \Delta \s{q_1}$, we find that the existence of such a classical correlation would require the variance of the momentum to be much smaller than the corresponding classical value, i.e.
    \beq
    \frac{\Delta_{\pi_ 1(dec)}}{|\pi_{1cl}|}=\frac{\sqrt{\s{C}_{11}\s{C}_{22}}}{|\s{C}_{12}|}\ll 1\,.
    \eeq
This condition is satisfied for both cases as can be seen in figure 3. The second case  exhibits an almost perfect degree of classical correlation by the time inflation ends and the oscillations of the heavier field have decayed. This behavior parallels that of the single field case, where the deviation from a perfect classical correlation of the field amplitude and the corresponding momentum is exponentially small \cite{Polarski:1995jg}. The first case also exhibits correlation between field amplitude and momentum but which is significantly lower; the variance in the momentum is about $10^{-4}$ of the classical momentum value. In principle, such an effect will tend to wash out the CMB acoustic peaks slightly. Although it seems unlikely that such a small relative variance in the momentum would have an observable effect on the CMB, it may be interesting to examine its consequences.

\begin{figure}[t]
    \includegraphics[width=7cm]{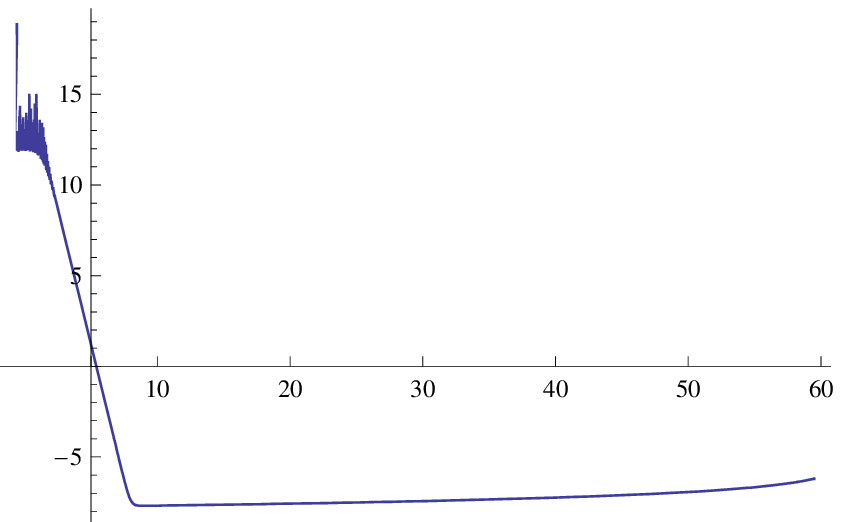}
    \includegraphics[width=7cm]{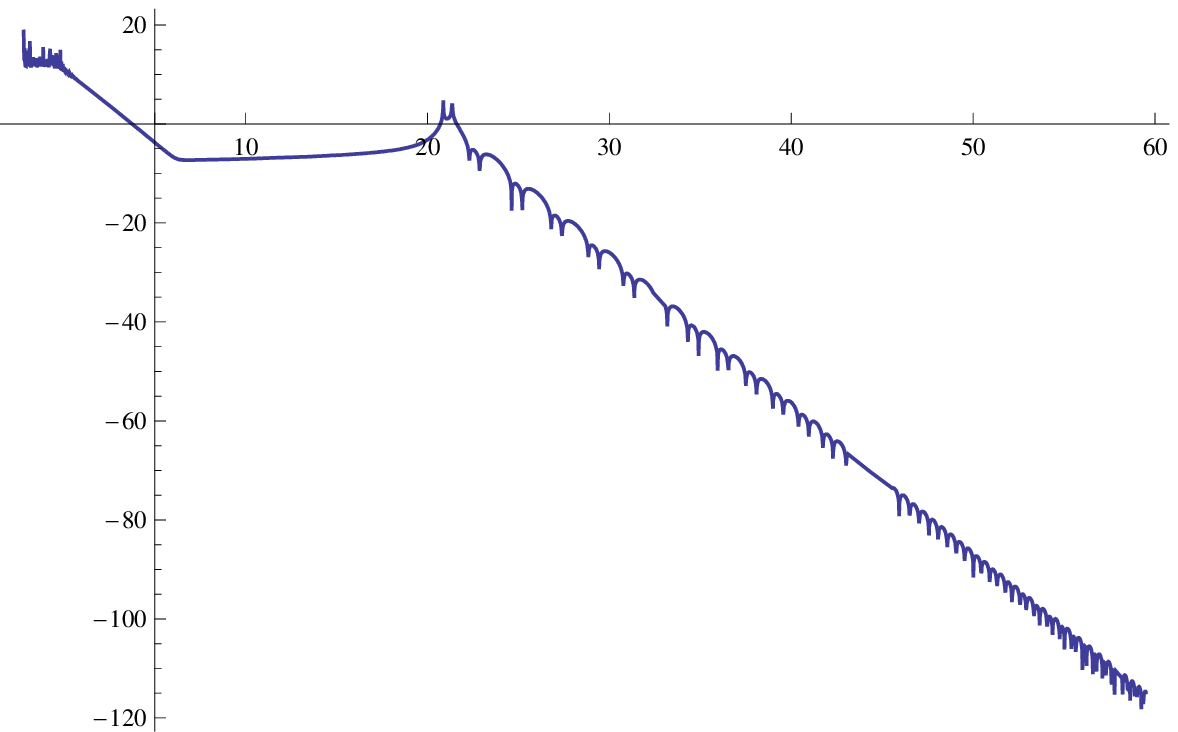}
    \caption{The classical correlation between field amplitude and momentum as quantified by the ratio $\ln \frac{\sqrt{\s{C}_{11}\s{C}_{22}}}{\s{C}_{12}}$ for $\mu=1.05 m$ (left) and $\mu=10 m$ (right).}
    \label{fig3}
    \end{figure}

%

\section{Discussion}

Let us close with a brief summary of the arguments presented in
this paper. We discussed the possible relevance of isocurvature
perturbations for the decoherence of cosmological curvature
perturbations generated during inflation. We have argued that
$\s{q}_1$ - $\Phi$ - $\s{q}_2$ can play for cosmological
perturbations the role of the triptych ``system'' - ``measuring
apparatus'' - ``environment'' which is central to considerations
of decoherence.  The crucial ingredient is the appearance of a non-zero
imaginary component $\s{B}_{12}$ on superhorizon scales which
results in $\s{C}_{11}/\s{C}_{22}\ll 1$ exponentially. We expect this phenomenon to be generic in all multi-field models since, in general, the imaginary parts are exponentially dominant over the real parts.
We explicitly checked a free two-field model where no explicit couplings
between the two fields were postulated. Nevertheless, adiabatic and
isocurvature perturbations are generically coupled through
gravitational interactions, implying that decoherence of this sort
will be relevant in general. Since this decoherence is very effective, we do
not expect the inclusion of non-linear interactions to affect this
result. Furthermore, even if the entropy -
and the momentum variance - increases, the strong classical correlation between field amplitude and momentum, responsible for the appearance of the CMB acoustic peaks, persists.


\section{Acknowledgments}
This research was supported by EU-RTN network MRTN-CT-2004-005104.


%

\end{document}